# Automatic detection and quantitative assessment of peculiar galaxy pairs in Sloan Digital Sky Survey

Lior Shamir[1,?], John Wallin[2]

1. Dept. of Comp. Sci., Lawrence Technological University
21000 W Ten Mile Rd., Southfield, MI 48075, USA
2. Dept. of Physics and Astronomy, Middle Tennessee State University 1301 E Main St,
Murfreesboro, TN 37130, USA

**ABSTRACT**

We applied computational tools for automatic detection of peculiar galaxy pairs. We first detected in SDSS DR7 ∼400,000 galaxy images with i magnitude <18 that had more than one point spread function, and then applied a machine learning algorithm that detected ∼26,000 galaxy images that had morphology similar to the morphology of galaxy mergers. That dataset was mined using a novelty detection algorithm, producing a short list of 500 most peculiar galaxies as quantitatively determined by the algorithm. Manual examination of these galaxies showed that while most of the galaxy pairs in the list were not necessarily peculiar, numerous unusual galaxy pairs were detected. In this paper we describe the protocol and computational tools used for the detection of peculiar mergers, and provide examples of peculiar galaxy pairs that were detected.

[?] Email: lshamir@mtu.edu

## 1 INTRODUCTION

Interactions between galaxies are associated with galaxy morphology (Springel & Hernquist 2005; Bower et al. 2006), quasars (Hopkins et al. 2005), enhanced rates of star formation (Di Matteo et al. 2007; Bridge et al. 2007), quasars (Hopkins et al. 2005), and activity in galactic nuclei (Springel et al. 2008). Numerous manually crafted catalogues and classification schemes of galaxy mergers have been proposed and published (Arp 1966; Struck 1999; Schombert, Wallin & Struck 1990; Vorontsov-Velyaminov 1959, 1977; Arp & Madore 1987). Cotini et al. (2013) developed and utilized a method for automatic detection of galaxy mergers, and studied galaxy merger population to show a link between mergers and galaxies with supermassive black holes.

More recent work on interacting systems has focused on pairs of galaxies with similar redshifts and small projected distances taken from the Sloan Digital Sky Survey. In these papers, the authors have systematically examined these close pairs for evidence of an increased star formation rate (Ellison et al. 2008), elevated nuclear activity (Ellison et al. 2010), and other measurable effects that might be associated with interaction. The confounding effect in these studies is the possibility of superpositions between galaxies in the same group. If the galaxies in a close pair have peculiar morphologies, there is a high confidence that there has been a recent interaction and the system is not just a chance superposition. However since there has been no clear objective way to define when a galaxy is "peculiar" (Naim & Lahav 1996), the objective criteria of velocity and projected distance has been the best way of analyzing a large sample of interacting pairs.

Early efforts to identify and catalog peculiar galaxies used photographic surveys. The difficulty of defining a set of objective criteria for peculiar galaxies can be illustrated by the classification schemes that have been used in these catalogs. The Catalog of Interacting Galaxies VorontsovVelyaminov (1959, 1977) contained 335 objects and placed peculiar galaxies into six primary categories: "HII-regions", "M51 type", "Nests", "Pairs", "Pseudo-Rings", "Comets", and "Enigmatic." The Arp Atlas of Peculiar Galaxies (Arp 1966) was a catalog of 332 peculiar and interacting systems. There were four primary overlapping categories for these objects including Spirals, Galaxies, E and E-like Galaxies, and Double Galaxies. Within

c

these groups, there were 37 other subgroups including "ring galaxies", "three-arm spirals", "galaxies with jets",' and "double galaxies with wind effects." Given the range of naming conventions and morphologies, the question of when a galaxy is "peculiar" has remained difficult to quantify. Catalogues of galaxy interactions have traditionally been produced by manual inspection of galaxy images by a few dedicated scientists. However as data sets have grown, the Galaxy Zoo project (and including Galaxy Zoo II and the broader Zooniverse efforts) have incorporated "citizen scientist" volunteers to classify galaxy morphologies from SDSS data (Lintott et al. 2008, 2011). Such manual analysis of galaxies using crowdsourcing was used for analyzing properties of merging galaxies (Darg et al. 2010; Casteels et al. 2013).

Aside from the questions about the completeness of these catalogs, the time needed to construct them is immense. Arp & Madore (1987) reported that it took ∼14 years to compile and produce their catalog of ∼6400 mergers in the southern hemisphere. In the era of robotic telescopes and digital sky surveys acquiring images of many billions of galaxies (Djorgovski et al. 2013; Borne 2013), manually crafted catalogues of mergers becomes impractical. While the morphology of most galaxy mergers is known, some galaxy mergers have peculiar morphology. Here we describe the detection of peculiar galaxy mergers by a computer algorithm mining galaxy images acquired by Sloan Digital Sky Survey.

## 2 METHOD

Galaxy mergers feature complex morphology that involves the shape of two or more interacting galaxies, as well as the distance and position of the galaxies in the system. Here we analyze images from from Sloan Digital Sky Survey (Schneider et al. 2003).

In the first step, we downloaded ∼ $3.7 \times 10^6$ objects identified by SDSS as galaxies (object type = 3) with i magnitude <18. Each galaxy image was of dimensionality of 120×120 pixels, downloaded directly using DR7 Catalog Archive Server (CAS) as jpg images, and converted to 8-bit TIFF format. The magnitude threshold is used to reduce the number of images to a smaller set of the brightest objects. Downloading all these images lasted 28 days.

After the images were downloaded, a preliminary test was applied to each image to filter possible artefacts. The preliminary test rejected images in which 80% or more of the pixels were brighter than 120, or images that 80% or more of the pixels were of the same colour (green, blue or red) after applying a fuzzy logic-based colour classification transform (Shamir 2006). The test was based on the SDSS colour mapping (Lupton et al. 2004), in which large swathes of a single colour are often signs of detector saturation, and can therefore be considered artefacts. After rejecting the artefacts, ∼ $3.2 \times 10^6$ objects were left.

Then, we applied an algorithm to determine whether a certain image has two neighboring objects, or that an object had more than one point spread function in it. The detection of two separate objects was detected by first applying the Otsu binary transform (Otsu 1979) to separate the foreground from the background pixels. The Otsu binary transform is performed by first computing the Otsu threshold (Otsu 1979). The Otsu method determines the threshold above which a pixel is considered a foreground pixel by iteratively testing each gray value, and computing the variance of the pixels dimmer than that value and the variance of the pixels brighter than the candidate threshold. The gray value that provides the minimum of the sum of the variances is the Otsu threshold (Otsu 1979). The Otsu threshold separates the foreground and background pixels regardless of linear mapping of the pixels intensity values.

The set of foreground pixels is then separated into foreground objects by counting the 4-connected objects (Shamir 2011a). The 4-connected objects are simply groups of foreground pixels such that each foreground pixel $I_{x,y}$ within the group $O$ satisfies the condition

$\forall I_{x,y} \in O \; \exists (I_{x,y+1} \in O | I_{x,y-1} \in O | I_{x+1,y} \in O | I_{x-1,y} \in O)$. That is, each foreground pixel in the group has at least one neighboring foreground pixel.

If more than one object is found, the image is flagged as a candidate for a galaxy merger. If only one object is found, the object is scanned for peaks using a point spread function detection algorithm (Shamir & Nemiroff 2005a,b), and if more than one peak is found the image is considered a potential galaxy merger. The peak detection code is part of the *Wolf* open source image analysis package (Shamir et al. 2006; Shamir 2012a). It should be noted that the same technique can also be used for automatic detection of recoiling supermassive black holes.

The separation of objects with more than one peak reduced the set of ∼ $3.2 \times 10^6$ images to ∼ $4.32 \times 10^5$ potential galaxy mergers. However, many of these images are not images of interacting galaxies. Figure 1 shows a few examples of objects classified as galaxies by SDSS pipeline and were also detected as potential galaxy mergers.

As the figure shows, images with two objects or with objects with two detected PSFs are not necessarily galaxy mergers. To find galaxy mergers we used the Wndchrm image analysis software tool (Shamir et al. 2008a; Shamir 2013b), which was originally developed for analysis of microscopy images (Shamir et al. 2008b, 2010a), but was also found informative for the analysis of galaxy images (Shamir 2009), and in particular for analyzing the morphology of galaxy mergers (Shamir et al. 2013a). Wndchrm works by first extracting a very large set of numerical image content descriptors for each image, so that each image is represented by a vector of 2883 numerical values. These content descriptors provide a comprehensive set that reflects the shape, colour, textures, fractals, polynomial decomposition of the image, and statistics of the pixel value distribution. These content descriptors are extracted from the raw images, but also from transforms of the images (e.g., FFT, Wavelet, Chebyshev, gradient), as well as combinations of transforms (e.g., FFT transform of the Wavelet transform). A detailed description of Wndchrm can be found in (Shamir et al. 2008a, 2010b, 2013a). Wndchrm performs colour analysis by using the RGB channels (Shamir et al. 2010b). That type of analysis might be less accurate than analyzing the FITS images of each colour channel separately, but it allows colour analysis without the need to download multiple FITS files for each celestial object, and therefore scales better when downloading and processing millions of galaxies.

Wndchrm was used by first manually classifying an initial set of 100 true galaxy mergers and another set of 100 images that are clearly not mergers. Then the image classifier was used to classify the galaxy images, and was inspected for classification errors. For each misclassified image, the image was added to the training set to improve the efficacy of the classification, leading to a training set of 500 samples. The training set was applied to the dataset of $\sim 4.32 \times 10^5$ galaxies, and reduced it to $\sim 2.1 \times 10^4$ images that were classified by Wndchrm as mergers. The image classification using Wndchrm takes ~45 seconds to classify a single galaxy image using a single core of Intel core-i7 processor, but since Wndchrm can be easily parallelized (Shamir et al. 2008a), a medium-sized cluster of 320-cores can process the galaxy images in less than one day.

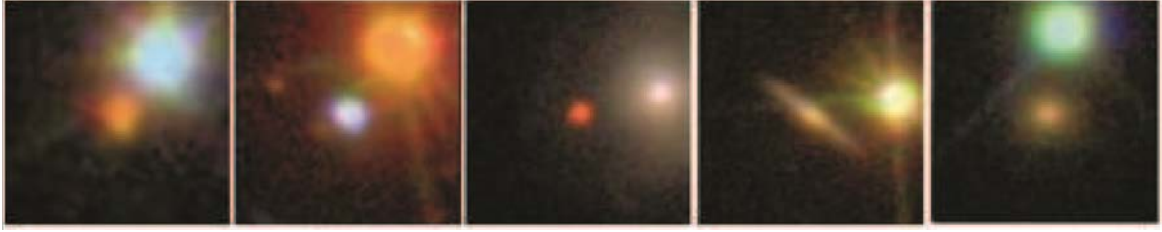

**Figure 1.** Objects detected as possible galaxy mergers

To find peculiar galaxy mergers, we then applied an algorithm for automatic detection of peculiar galaxies (Shamir 2012b) that works by weighting the image content descriptors computed by Wndchrm such that the weights are determined using the variance of the values in the training set, and then measuring the weighted Euclidean distance between each image in the test set and the "typical" image in the training set. The algorithm is based on a peculiar image detection algorithm (Shamir 2013b), and was applied to galaxy images as described in detail in (Shamir 2012b). Experimental results and a detailed description about the peculiar image detection algorithm is provided in (Shamir 2012b, 2013b). From the output of the peculiar image detection algorithm we took the top 500 images and inspected them manually.

## 3 RESULTS

Automatic detection of peculiar galaxies is a complex task, and it is expected that such algorithms will have a certain degree of noise. Due to the noise, many of the 500 galaxy pairs detected by the algorithm were not peculiar, and some also contained artefacts that were not filtered in the previous stages. However, among the shortlist of galaxy pairs many images of peculiar galaxy pairs were found. Although the algorithm had to rely on a last step of manual inspection, it reduced a list of $\sim 3.7 \times 10^6$ images of celestial objects into a manageable list of 500 candidate objects. From that list, artefacts and non-peculiar galaxies were removed by manual examination of the galaxy pairs, and the most peculiar objects were selected manually by the authors.

Figure 2 shows the image, DR7 object ID, and celestial coordinates of some of the objects detected by the algorithm. As mentioned above, most celestial objects in the list of 500 objects were not peculiar or did not have clear morphology, and are therefore not included in this paper. The list of 500 celestial objects is available as supplementary on-line material of this paper.

Figure 2. Examples of the galaxies identified by the method. The identification number below each galaxy image is the SDSS DR7 object identification number. The number above each image is an identifier by which the galaxy pair is identified in the paper

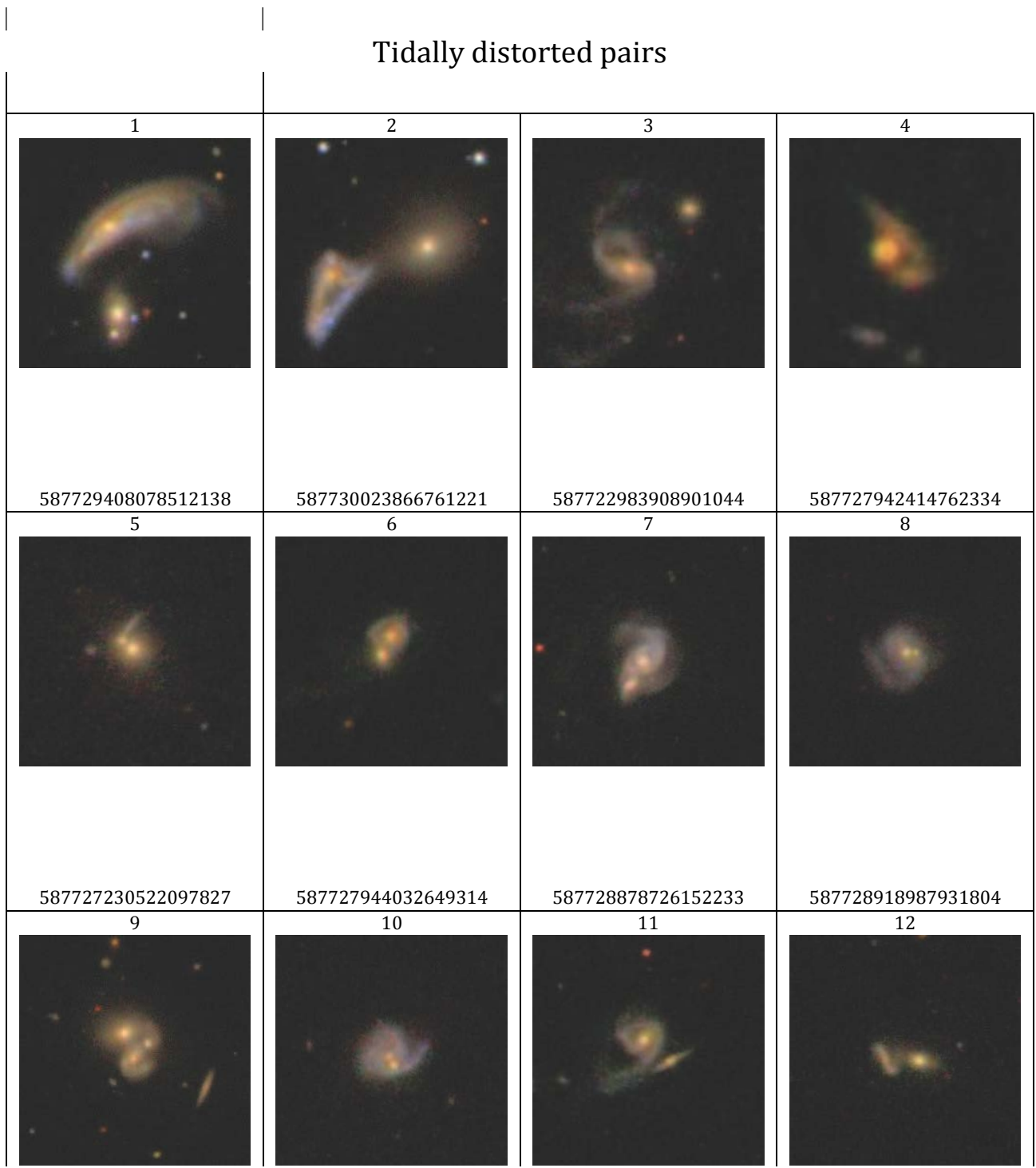

| 13 | 14 | 15 | 16 |
|---|---|---|---|
| 587729772070633570 | 587739156573585582 | 587741710474870997 | 587742772949549151 |
| 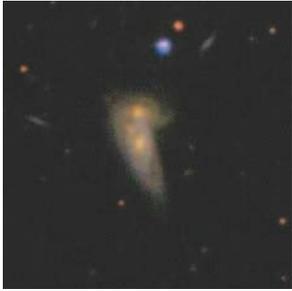 | 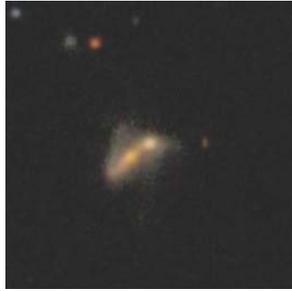 | 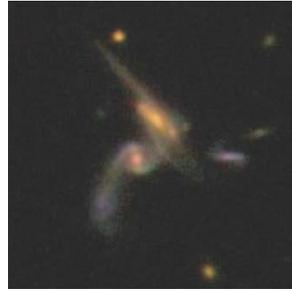 | 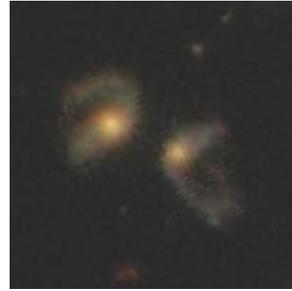 |
| 587744875857642337 | 587746236302360825 | 588015509281833181 | 587736942524629287 |
| 17 | 18 | 19 | 20 |
| 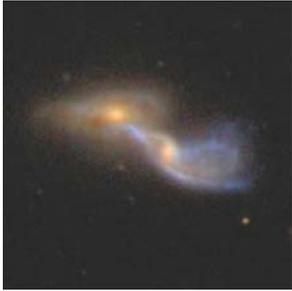 | 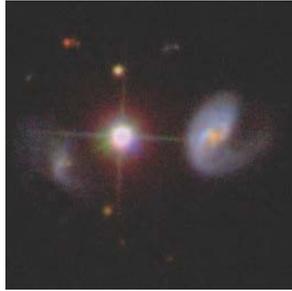 | 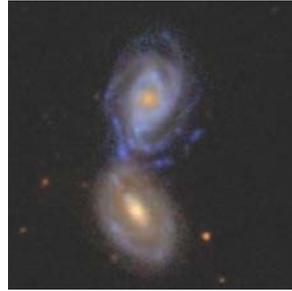 | 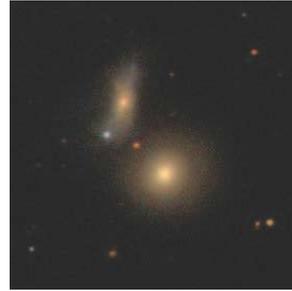 |
| 588298664650145819 | 587724232641937419 | 587731499185864817 | 587736974735704203 |

## Collisional Ring Galaxies

| 21 | 22 | 23 | 24 |
|---|---|---|---|
| 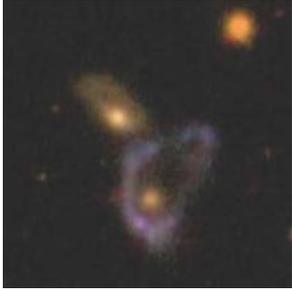 | 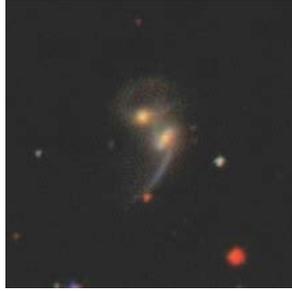 | 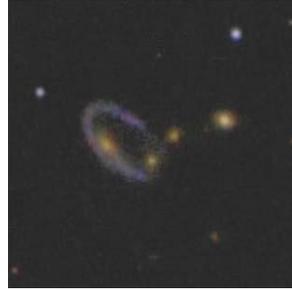 | 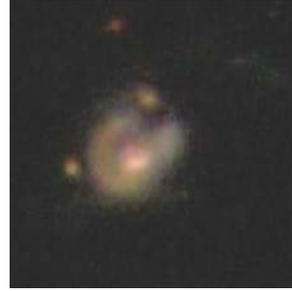 |
| 587730774407840452 | 587742631737229751 | 587730845814751853 | 587728308567015452 |

| 25 | 26 | 27 | 28 |
|---|---|---|---|
| 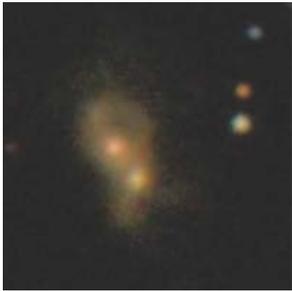 | 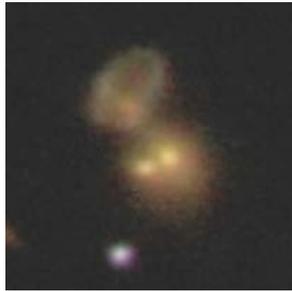 | 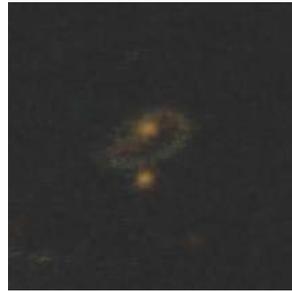 | 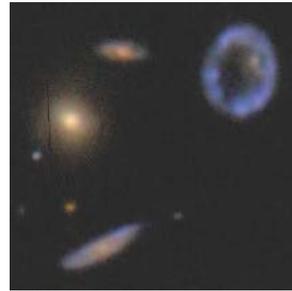 |
| 587729233595859458 | 587736976890134917 | 587731187810238739 | 587725550139277460 |

## Blue Galaxies with Unusual Morphologies

| 29 | 30 | 31 | 32 |
|---|---|---|---|
| 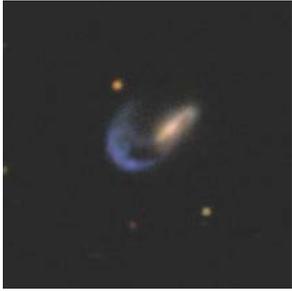 | 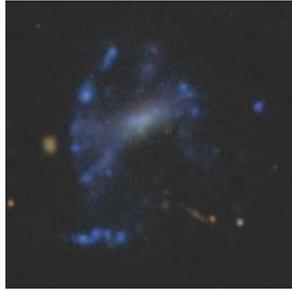 | 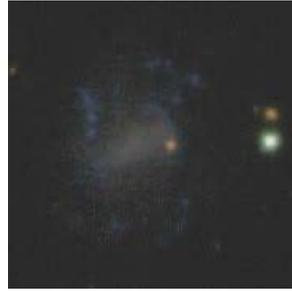 | 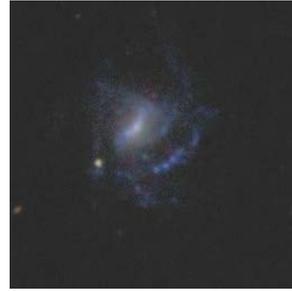 |
| 587739305294626830 | 587731500262948921 | 587741533327458358 | 587739407295905821 |
| 33 | 34 | 35 | 36 |
| 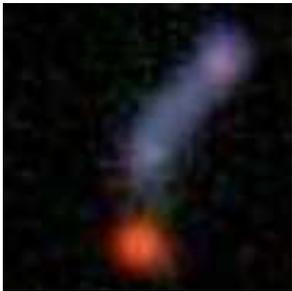 | 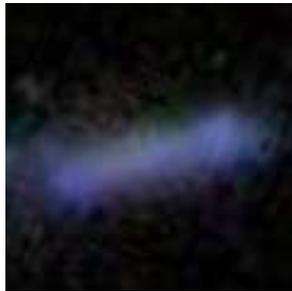 | 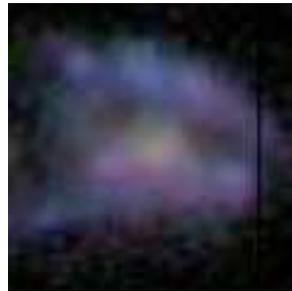 | 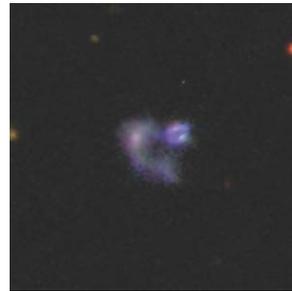 |
| 587729772072861802 | 587732484351590536 | 588848900451008597 | 587725775608086591 |

## Galaxies with Embedded Point Sources

| 37 | 38 | 39 | 40 |
|---|---|---|---|
| 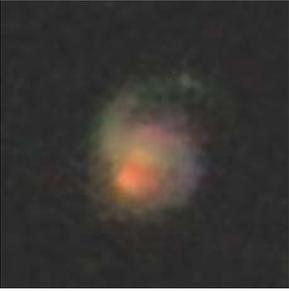 | 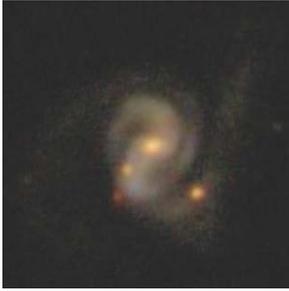 | 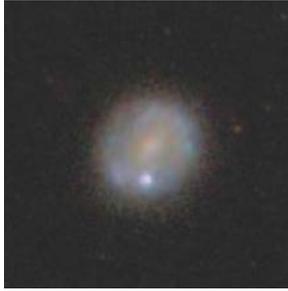 | 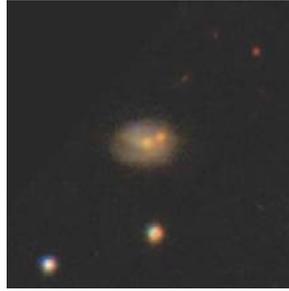 |
| 587729233591861508 | 587728676861182142 | 587726102561161239 | 587733434070860127 |
| 41 | 42 | 43 | 44 |
| 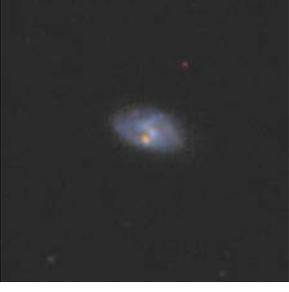 | 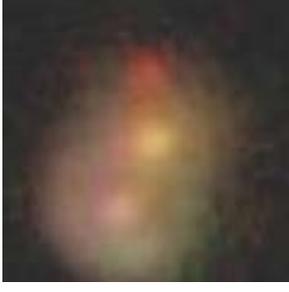 | 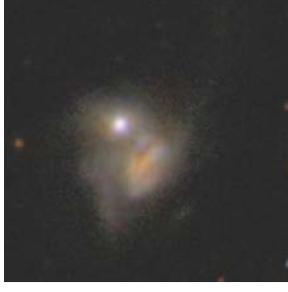 | 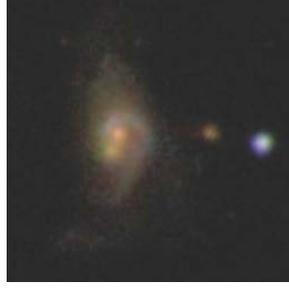 |
| 587742015424233553 | 587724232110440585 | 587729233051648085 | 587729653430222900 |
| 45 | 46 | 47 | 48 |
| 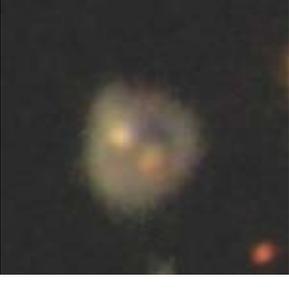 | 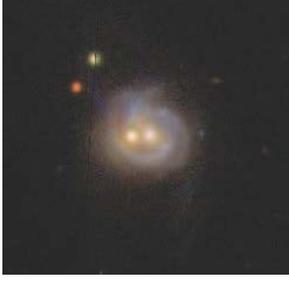 | 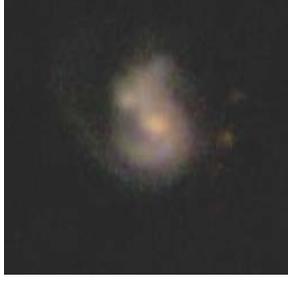 | 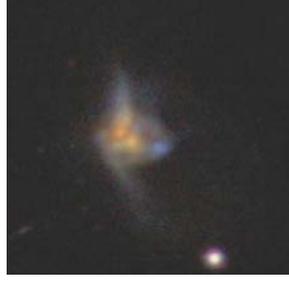 |
| 587733410446180362 | 587742013279502419 | 587740522398089357 | 587732134852427839 |

## Edge-on Galaxies and Linear Features

| 49 | 50 | 51 | 52 |

| 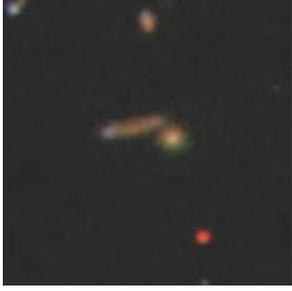 | 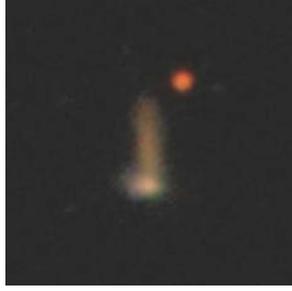 | 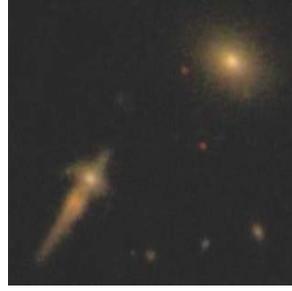 | 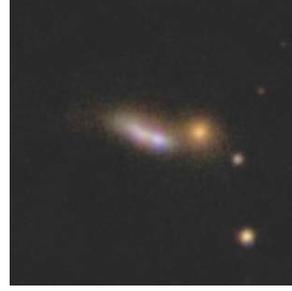 |
|---|---|---|---|
| 587729776907452608 | 588011219678527613 | 587731681724531032 | 587740552454340671 |
| 53 | 54 | 55 | 56 |
| 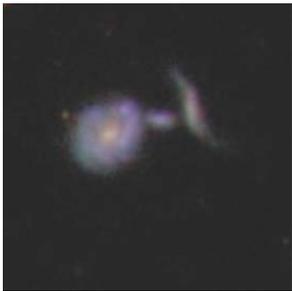 | 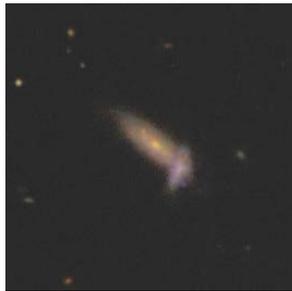 | 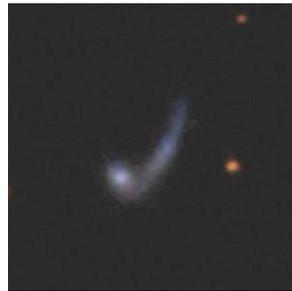 | 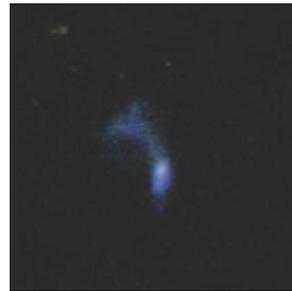 |
| 587724232104280165 | 587742863676473541 | 587732702867095573 | 588017704018313247 |
| 57 | 58 | 59 | 60 |
| 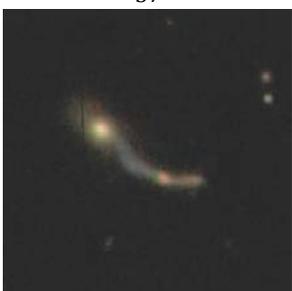 | 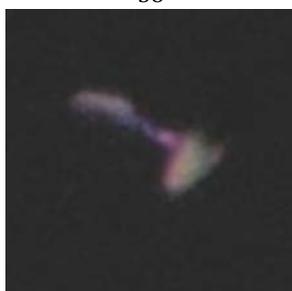 | 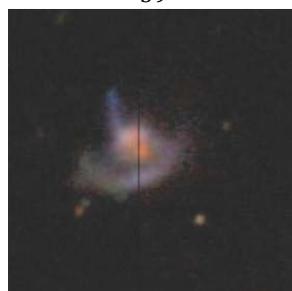 | 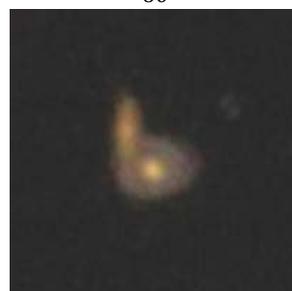 |
| 587741602571223188 | 587739380459700413 | 587726015607341151 | 587739721384264013 |

Table 1: Tabular data on the example galaxies.

| No. | SDSS ID | RA (degrees) | Dec (degrees) | $z$ | Descriptions and cross identifications |
|---|---|---|---|---|---|
| 1 | 587729408078512138 | 249.55762 | 41.93106 | 0.028 | Tidally distorted interacting pair Arp 125/UGC 10491 |
| 2 | 587730023866761221 | 222.85136042 | 6.80141814 | 0.035 | Tidally distorted interacting pair [RC2] A1488+07B/CGCG 048-028N |
| 3 | 587722983908901044 | 228.80236789 | 0.48465685 | - | Tidally distorted interacting pair |
| 4 | 587727942414762334 | 130.41641554 | 0.47482966 | - | Tidally distorted interacting pair |
| 5 | 587727230522097827 | 24.57851606 | -9.53761984 | 0.105 | Tidally distorted interacting pair |
| 6 | 587727944032649314 | 146.96981692 | 2.2030615 | 0.100 | Tidally distorted interacting pair |
| 7 | 587728878726152233 | 146.07461364 | 2.82742562 | 0.061 | Tidally distorted interacting pair |
| 8 | 587728918987931804 | 228.93245046 | 57.32576843 | 0.069 | Tidally distorted interacting pair |
| 9 | 587729772070633570 | 204.18173946 | -3.49915134 | 0.053 | Tidally distorted interacting pair CGCG 017-034 |
| 10 | 587739156573585582 | 133.73038433 | 24.57972348 | -0.000 | Late stage merger + second nucleus |
| 11 | 587741710474870997 | 129.7030717 | 17.77376393 | 0.118 | Tidally distorted interacting pair |
| 12 | 587742772949549151 | 183.55630868 | 16.39304707 | 0.106 | Tidally distorted interacting pair |
| 13 | 587744875857642337 | 118.40246117 | 9.39599884 | - | Tidally distorted interacting pair CGCG 058-063 |
| 14 | 587746236302360825 | 213.24517723 | -15.64564401 | - | Tidally distorted interacting pair |
| 15 | 588015509281833181 | 28.43139742 | 0.18638246 | 0.082 | Tidally distorted interacting pair |
| 16 | 587736942524629287 | 230.40263074 | 31.31348877 | 0.107 | Tidally distorted interacting pair |
| 17 | 588298664650145819 | 194.20114032 | 48.29557185 | 0.028 | A strongly interacting pair of disk galaxies NGC 4837/I Zw 046 |
| 18 | 587724232641937419 | 20.01095866 | 14.36176367 | 0.031 | Pair of tidally distorted galaxies next to a bright star CGCG 058-063 |
| 19 | 587731499185864817 | 155.94389483 | 53.10296621 | 0.031 | Close pair of spirals - some evidence of tidal distortion UGC 05615/VV 312/CGCG 1020.6+5321 |
| 20 | 587736974735704203 | 231.39225706 | 26.55545497 | 0.034 | Spiral + Elliptical interacting pair CGCG 165-053 |
| 21 | 587730774407840452 | 318.5274773 | 10.60878036 | 0.089 | Double ring galaxy |
| 22 | 587742631737229751 | 257.37340763 | 42.53970985 | - | Tidally distorted interacting pair |
| 23 | 587730845814751853 | 315.55380799 | -1.19879418 | 0.100 | Collisional ring galaxy |
| 24 | 587728308567015452 | 172.03453114 | 2.3942382 | - | Apparent ring galaxy? |
| 25 | 587729233595859458 | 260.34558059 | 33.72469744 | - | Apparent ring galaxy and intruder |
| 26 | 587736976890134917 | 247.3840953 | 20.33012694 | 0.092 | Apparent ring galaxy and pair of ellipticals |
| 27 | 587731187810238739 | 350.15009605 | 1.18186531 | - | Collisional ring |
| 28 | 587725550139277460 | 188.02672417 | 66.40332813 | 0.048 | Collisional ring UGC 07683/VV 788, VII Zw 466/ CGCG 315-043/[RC2] A1229+66B |
| 29 | 587739305294626830 | 195.73997132 | 35.66516 | 0.037 | Incomplete ring galaxy |
| 30 | 587731500262948921 | 167.90750782 | 56.51715493 | 0.010 | Unusual irregular galaxy |
| 31 | 587741533327458358 | 184.20861132 | 30.27156024 | 0.013 | Irregular galaxy |
| 32 | 587739407295905821 | 126.70406448 | 20.36484288 | 0.025 | Irregular galaxy IC 2373/UGC 04409/CGCG 119-100 |
| 33 | 587729772072861802 | 209.30600585 | -3.36564749 | - | An irregular blue galaxy next to reddish star |
| 34 | 587732484351590536 | 155.00631865 | 46.59967906 | 0.030 | An isolated irregular blue galaxy |
| 35 | 588848900451008597 | 183.27620792 | 0.212918 | 0.096 | An irregular blue galaxy |
| 36 | 587725775608086591 | 121.73547563 | 48.51907092 | 0.078 | One-armed spiral and companion |
| 37 | 587729233591861508 | 254.77478876 | 41.80435125 | - | Superposition between a barred spiral and a star? |
| 38 | 587728676861182142 | 203.67710948 | 62.57444007 | 0.076 | Multiarm barred spiral |
| 39 | 587726102561161239 | 222.7289759 | 4.94891811 | 0.014 | Spiral with bright star in its disk |
| 40 | 587733434070860127 | 254.71927935 | 28.30197041 | - | Red with star |
| 41 | 587742015424233553 | 167.11388932 | 22.61855803 | 0.022 | Blue spiral with star |
| 42 | 587724232110440585 | 32.64724334 | 12.91822181 | 0.1 | Merging pair |
| 43 | 587729233051648085 | 248.46782084 | 47.99532843 | 0.035 | Distorted galaxy with nearby star CGCG 251-028 |
| 44 | 587729653430222900 | 261.09832143 | 25.60862749 | - | Single arm spiral and companion |
| 45 | 587733410446180362 | 213.91507664 | 50.71345968 | 0.049 | Single arm spiral and companion |
| 46 | 587742013279502419 | 174.12239999 | 21.59607456 | 0.030 | Spiral galaxy with a second nuclear source NGC 3758 |

| 47 | 587740522398089357 | 8.89698672 | 23.76797938 | - | Close interacting pair with tidal distortions |
| 48 | 587732134852427839 | 195.92079205 | 51.49684627 | 0.038 | Merging pairs |
| 49 | 587729776907452608 | 214.6336745 | -2.61358823 | - | Close pair |
| 50 | 588011219678527613 | 229.9230259 | 54.82464843 | 0.115 | Close pair |
| 51 | 587731681724531032 | 122.52209047 | 35.16653732 | 0.087 | Edge-on interaction of to spirals |
| 52 | 587740552454340671 | 54.69581337 | 15.54806086 | - | Close pair |
| 53 | 587724232104280165 | 18.14808494 | 14.01249256 | 0.053 | Close group of three galaxies |
| 54 | 587742863676473541 | 181.80239643 | 16.96934309 | 0.072 | Possible superposition? CGCG 098-060 |
| 55 | 587732702867095573 | 162.928461 | 7.2946855 | 0.023 | Close pair |
| 56 | 588017704018313247 | 212.82335434 | 11.3211844 | 0.028 | Tidally distorted galaxy |
| 57 | 587741602571223188 | 190.90262203 | 27.89193356 | 0.083 | Close pair |
| 58 | 587739380459700413 | 239.01810342 | 21.8666069 | 0.085 | Close pair |
| 59 | 587726015607341151 | 183.44198571 | 2.81152618 | 0.073 | Late stage |
| 60 | 587739721384264013 | 242.60781765 | 17.76026039 | 0.129 | Possible superposition between two spiral galaxies |

## 3.1 Characteristics of Targets

In Figure 2 we present a set of sixty images of galaxies that were detected using our method. Some of these images are clearly strongly interacting systems, while others have unusual morphologies. It would be impossible to represent all of the types of galaxies found by the algorithm, but these groupings help inform the types of features that the algorithm finds unusual enough to be flagged.

### 3.1.1 Tidally Distorted Pairs

Galaxies in this category are close pairs of interacting systems with obvious tidal distortion. The first set of twenty galaxies from Table 1 shows some examples of these kinds of morphologies. In images such as galaxies 1 and 2, we can see highly distorted spiral galaxies. These systems are clearly in the late stages of a merger, but rather shortly after the close approach between the galaxies in these images. The galaxies in images 5 through 9 also have strong tidal features, but seem to be examples of older interactions. In all of the examples of tidally distorted pairs, the galaxies do not follow the patterns typically associated with either spiral or elliptical galaxies, and there are two identifiable progenitors in the system with at least one showing clear signs of tidal disruption.

### 3.1.2 Collisional Ring Galaxies

Galaxies 21 through 28 in Table 1 are examples of collisional ring galaxies (Appleton & Struck 1996). Galaxies 21, 22, 23, 27, and 28 have very bluish colours in their rings suggesting the enhanced rates of star formation commonly seen in these systems. Galaxies 24, 25, and 26 have less well defined ring structures with less bluish colours. It is possible that these systems may have progenitors with less gas resulting in very low rates of new star formation. Other ring galaxies such as in AM1724-622 exhibit similar behavior (Wallin & Struck-Marcell 1994). In galaxy pair 28 spectroscopic redshift is available for the two blue galaxies (the ring galaxy and the galaxy in the lower left part of the field), and for both galaxies the redshift is 0.048.

### 3.1.3 Blue Galaxies with Unusual Morphologies

Galaxies 29 through 36 have strong blue colours and unusual morphologies. Galaxy 29 is an interacting pair that underwent an interaction similar to those that formed the collisional ring galaxies in the previous section. However, galaxies 30, 31, and 32 seems to be blue spirals with irregular structures. Galaxies 33 and 34 have no obvious features such as a nucleus or spiral arms. They are clearly elongated galaxies with blue colours.

### 3.1.4 Galaxies with Embedded Point Sources

Galaxies 37 through 47 have secondary point sources in their disks or envelopes. In many cases, such as galaxy pairs 42, 44, 45, 46, 47 and 48, there is a bulge-like concentration in the system. These are likely late-stage mergers. For galaxies 37, 38, 39, 40, 41, and 43, there is a clear secondary point source in the disk but its origin is less clear. In some cases this may be an embedded supernova in the galaxy or perhaps a chance superposition with a foreground star.

### 3.1.5 Edge-on Galaxies and Linear Features

Galaxies 49 through 60 are edge-on galaxies or galaxies with long, thin features. These elements are not technically linear, but rather thin extended features that are not typically found in galaxies. In some cases such as 49, 50, 51, and 52, these may be simple super positions of an edge-on disk galaxy with a second galaxy. There are more obvious tidal features in some of the images such as 55, 56, 57 and 58. These systems seem to have connecting bridges and tidal tails on one of the galaxies. Galaxy 59 appears to be a latestage merger.

## 3.2 Discussion

For most of the galaxy pairs, the categorization could be into other groups. For example, galaxy pair 60 could easily be put into the category of tidally distorted groups. Additional categories could also be created to capture some of the subgroups in these systems. It would, for example, be tempting to create separate categories of early- and late-stage mergers. However, the categories and examples we have chosen are designed to illustrate systems with common visual elements that the algorithm is likely to find unusual, and the reasons why they were flagged as morphologically peculiar.

It is also important to point out that some of these galaxies have been seen before. Galaxy 1, for example, is Arp 125. Several other of the examples appear in older catalogs of galaxies and clusters.

Given the rich variety of galaxy types, it may be possible to perform additional automated classification of the images into different subcategories if a sufficiently large sample is used for training and testing. Unsupervised learning might make it possible to better understand the features that the algorithm finds peculiar. The analysis performed in this paper was done by an image analysis method that uses very many numerical image content descriptors, and the high dimensionality of the analysis makes it highly difficult to conceptualize the criteria by which a certain combination of feature values is considered peculiar.

It can be reasonably assumed that some of the detected celestial objects may be pairs of galaxies that have no gravitational interaction, but happen to be in the same field due to super-positioning (Karachentsev 1985, 1990). Since in most cases spectroscopic z is not available for both objects, it is possible to use the photometric z to obtain rough estimation whether the two objects are part of the same system or adjacent only in projection. Having accurate velocity measurements of these objects would help remove this ambiguity, but not eliminate it completely, especially for z<0.1, for which SDSS photometric redshift is less accurate. Galaxy pairs 6, 7, 16, 28, 29, 32, 57 and 59 all have the same spectroscopic z for both objects. In the case that spectroscopic z is not available, we compared the photometric redshift, and the detected galaxy pairs have similar photometric redshifts.

For instance, galaxy pair 60 has spectroscopic redshift of 0.129, and the nearby galaxy has photometric redshift of 0.134. More importantly, a larger sample of interacting galaxies with clear tidal distortions can be used to train the algorithm further to identify systems that are unambiguously interacting.

## 4 CONCLUSION

Autonomous digital sky surveys have been generating vast pipelines of astronomical images, leading to big astronomical databases. This form of astronomical data collection cannot rely solely on manual analysis, and requires algorithms and computer methods that can process these data and transform them into smaller and well-defined datasets that can be effectively used by humans.

Here we show how an automatic method can mine through a large dataset of $\sim 3.7 \times 10^6$ galaxy images and reduce them to a list of 500 images, containing many peculiar galaxy mergers. Detecting these peculiar mergers manually in a dataset of almost four million celestial objects is very difficult to perform manually, and can be considered nearly impractical without using automatic data analysis tools. Future digital sky surveys such as LSST will provide clear morphology of billions of celestial objects, magnifying the problem of detection of peculiar galaxies by an order of magnitude and making manual detection of such objects virtually impossible. Reduction of the data to much shorter lists as was demonstrated in this work using SDSS data will make the detection of peculiar galaxies practical, or will allow the use of citizen science (Lintott et al. 2008) to analyze such future databases.

The source code for the automatic detection of peculiar images is publicly available, and can be downloaded at http://vfacstaff.ltu.edu/lshamir/downloads/chloe.
The source code for the *Wndchrm* method (Shamir 2013b) that computed the numerical image content descriptors can be accessed at http://vfacstaff.ltu.edu/lshamir/downloads/ImageClassifier.

## 5 ACKNOWLEDGMENTS

We would like to thank the reviewer, William Keel, for the insightful comments that helped to improve the manuscript. The computing cluster that was used to process the data was supported by NSF grants number 1157162.